\documentclass[final]{aipproc}
\layoutstyle{6x9}

\newcommand{\be}{\begin{equation}}
\newcommand{\ee}{\end{equation}}

\newcommand{\bef}{\begin{figure}}
\newcommand{\eef}{\end{figure}}

\newcommand{\bea}{\begin{eqnarray}}    
\newcommand{\eea}{\end{eqnarray}}

\def\spose#1{\hbox to 0pt{#1\hss}}      
\def\ltapprox{\mathrel{\spose{\lower 3pt\hbox{$\mathchar"218$}}      
\raise 2.0pt\hbox{$\mathchar"13C$}}}      
\def\gtapprox{\mathrel{\spose{\lower 3pt\hbox{$\mathchar"218$}}      
\raise 2.0pt\hbox{$\mathchar"13E$}}}      
\def\inapprox{\mathrel{\spose{\lower 3pt\hbox{$\mathchar"218$}}      
\raise 2.0pt\hbox{$\mathchar"232$}}}

\begin{document}

\title{Gravitational clustering: an overview}



\author{Francesco Sylos Labini}{
  address={``E. Fermi'' Center, Via Panisperna 89 A, Compendio del
Viminale, I-00184 Rome, Italy,\\ \& ISC-CNR, Via dei Taurini 19,
I-00185 Rome, Italy.}
}

\begin{abstract}
We discuss the differences and analogies of gravitational clustering
in finite and infinite systems. The process of collective, or violent,
relaxation leading to the formation of quasi-stationary states is one
of the distinguished features in the dynamics of self-gravitating
systems. This occurs, in different conditions, both in a finite than
in an infinite system, the latter embedded in a static or in an
expanding background. We then discuss, by considering some simple and
paradigmatic examples, the problems related to the definition of a
mean-field approach to gravitational clustering, focusing on role of
discrete fluctuations. The effect of these fluctuations is a basic
issue to be clarified to establish the range of scales and times in
which a collision-less approximation may describe the evolution of a
self-gravitating system and for the theoretical modeling of the
non-linear phase.
\end{abstract}

\maketitle


\section{Introduction}

As discussed in various papers in this volume (see
e.g.\cite{intro,campa}) equilibrium properties of long-range interacting
systems require a non-trivial analysis as standard thermodynamics
techniques do not simply apply when dealing with pair-interactions
decaying with sufficiently small exponents.  Many interesting and
unsolved problems lie in the out-of-equilibrium dynamics of systems
with long-range interaction about which very little is known from a
theoretical point of view (see also
\cite{chavanis} in this volume).
The understanding of the thermodynamics and dynamics of systems of
particles interacting only through their mutual Newtonian self-gravity
is of fundamental importance in cosmology and astrophysics. It
encompasses the range of physical scales relevant to the formation of
the largest structures in the Universe, down to those relevant to
stellar dynamics. The statistical mechanics of systems dominated  by
gravity has been studied and applied in many different contexts in
astrophysics and cosmology (see
e.g. 
\cite{lyndebell,pad_physrep,chavanis,chandra,chandra_revmodphy,
pad_book,pad_dtslri,saslaw1,saslaw2,binney,peebles,thierry}):
for example in the studies of globular clusters, galaxies and the
clustering in the expanding universe.

While systems with short range interactions can be usually studied
through laboratory experiments, gravitational systems can only be
observed in astrophysical contexts. Alternatively one may set up
numerical experiments which then represent the unique instrument to
study the dynamics of gravitational clustering. In this respect the
astrophysicist's perspective is usually to model some intricate
realistic systems, such as stellar or galaxy systems, having the aim
of understanding a specific set of observations.  For example in the
cosmological context one uses very complicated initial conditions
(described by a large number of parameters) and needs a certain number
of important assumptions, from the way the universe expands to the
amount and type of dark matter which dominates the dynamics on the
relevant scales. This is so because, by studying gravitational
clustering, one would like to understand the relations between some
important observations of the cosmos. For example the studies of the
cosmic microwave background radiations provide with the information
about the initial conditions of the matter density field. The large
scale geometrical properties of the universe are deduced, for example,
through the measurements of the supernovae magnitude-redshift
relation. Galaxy redshift surveys map the present-day matter
distribution. The estimations of the mass-to-light ratio of
astrophysical objects is ultimately related to the abundance of dark
matter. The task of the model of cosmological structure formation is
thus to build a unified and coherent picture to explain these (and
other) observations of the universe at the largest scales
\cite{peacock}.

In statistical physics the problem of the evolution of
self-gravitating classical bodies has been relatively neglected,
primarily because of the intrinsic difficulties associated with the
attractive long-range nature of gravity and its singular behavior at
vanishing separation.  When approaching the problem of gravitational
clustering in the context of statistical mechanics it is natural to
start by reducing as much as possible the complexity of the analogous
cosmological or astrophysical problem. For example, in order to focus
on the essential aspects of the problem one may study gravitational
clustering without the expansion of the universe, and starting from
particularly simple initial conditions. With respect to the motivation
from cosmology/astrophysics, there is of course a risk: in simplifying
we may loose some essential elements which change the nature of
gravitational clustering. Even it were, it seems unlikely that we will
not learn something about the more complex cosmological/astrophysical
situations  in addressing slightly different and simplified problems.

A fundamental distinction has to be made between finite and infinite
systems.  They have in common that the gravitational force on a
arbitrary point has contributions coming from all scales in the
system. However they differ for the fact that in the case of the
finite system there is a mean field force generated by the system as a
whole, which is related to its internal symmetries (e.g., spherical
symmetry) and which eventually will give rise to a global collapse of
the entire system. In an infinite space, in which the initial
fluctuations are non-zero and finite at all scales, the collapse of
larger and larger scales will continue ad infinitum: being no
geometric center there will not be a global collapse of the entire
system. The mean field dynamics is driven by system's fluctuations
which determine the gravitational force at different spatial scales.
The collapse occurring on larger and larger scales will clearly happen 
at different times but it is characterized by the unique time-scale in
the system that is
\be 
\label{tau} 
\tau \sim \sqrt{G\rho_0}^{-1} \,.  
\ee 
This is in general the typical characteristic time scale of any
(finite or infinite) gravitational system with average mass density
$\rho_0$. For instance, as we discuss below, this is the time scale
predicted by the self-gravitating fluid approximation.
  
The infinite system can therefore never reach a time independent
state, and in particular it will never reach a thermodynamic
equilibrium. Although so different, the finite and the infinite
systems share some subtle and important analogies which we briefly
discuss in what follows.  We firstly discuss the general difficulties
related to the long-range character of gravity and the usual way to
make a mean-field approximation. Then we consider two basic examples
of a finite and of an infinite system: the simplest example of a
finite system is represented by an initially isolated spherical
distribution of randomly distributed points (i.e. a Poisson
distribution) --- this will be also discussed in the contribution by
Morikawa in this book \cite{morikawa}.  Analogously the simplest
example of an infinite system is a Poisson distribution in an infinite
space. For this second case one may consider that the space background
is static (as Joyce in his contribution in this book \cite{joyce}) or
it is expanding (as Saslaw in his contribution \cite{saslaw}). We will
discuss these two cases briefly, outlining the analogies and the
differences. In the conclusions we try to point out which are the main
problems of the field.

\section{Gravitational relaxation and mean field models} 
\label{force}

The typical feature of short-range interactions system is that the
force between two particles (e.g. molecules of an ordinary gas) is
strong when they are very close to each other, i.e. when they repeal
each other strongly, and on large distances no force is exerted
between two particles. In this way particles most of time move at
nearly constant velocity, and then they are subject to violent and
short lived accelerations when they collide with one another. In this
situation, the typical mechanism for relaxation is that due to
two-body collisions. For self-gravitating particles the situation is
different because gravity is divergent at vanishing separation and it
is long range. Because of the former property the actual force on a
given particle maybe more or less strongly influenced by fluctuations
in its local neighborhood, while the latter property implies that a
particle is coupled with all other particles at all scales in the
system.  Thus the studies of a self-gravitating systems have to
consider that many different range of scales, in principle all, maybe
almost equally important for the analysis of the dynamical properties.

For this reason, one the principal problems of stellar dynamics
\cite{chandra,chandra_revmodphy} is concerned with the analysis of the nature
of the gravitational force acting, for example, on a star which is
member of a stellar system. In a general way we may broadly
distinguish between the influence of the system as a whole
$\vec{F}_{sm}$ and of the immediate neighborhoods $\vec{F}_{f}$.  The
former will be smoothly varying function of position and time while
the latter will be subject to relatively rapid fluctuations:
\[ 
\vec{F} = \vec{F}_{sm} + \vec{F}_{f} \;.
\] 
The second term describes  the fluctuations that are related to the
underlying statistical properties of the particle distribution: their
effects can be evaluated, under certain assumptions, in a stochastic
sense.

The problem is general is to establish a criterion to quantitatively
determine the circumstances under which
\be
\label{fapp}
\vec{F} \approx \vec{F}_{sm} \;,
\ee
and thus $\vec{F}_{f} \approx 0$, i.e. under which the system can be
treated by the mean field approximation. This consists in neglecting
the local interaction and to consider the global gravitational field
of the system.  This is not an easy task for the general case. For
example in the simplest situation of an isolated self-gravitating
system in virial equilibrium it is possible to estimate a criterion
for using the approximation provided with Eq.\ref{fapp}, by considering
the role of two-body scattering
\cite{chandra,binney,saslaw2}. Whether this can be used for the more
general case of an out-of-equilibrium system is an open question
\cite{binneyknebe,diemond,power02}.

As mentioned above, the mean field approximation consists in neglecting
the physics on  small scales in the system and for this reason it
usually describes a collision-less system. We briefly recall the main
steps to obtain such an
approximation~\cite{binney,pad_physrep,saslaw2}.  Let us divide the
phase space of a given system into a large number of cells in such a
way that (i) each cell is large enough to contain a macroscopic number
of particles (ii) each cell is small enough so that all particles in
the cell can be assumed to have the same average characteristic
properties of the cell. Thus the size of the cell should be large
enough to satisfy (i) but not so large to violate (ii). If such an
intermediate size can be defined at a generic time $t$ it is possible
to define smooth functions, as for example the one-point distribution
function. As the collapse will involve progressively larger and larger
scales this approximation may break down at a certain time, when
non-linear objects will be formed on the scale of the cell.

In the approximations above discussed the one-point distribution
function will satisfy the Boltzmann equation. By neglecting the
collision terms one may obtain the collision-less Boltzmann equation
or Vlasov equation which is the basic tool to study the evolution of
self-gravitating systems. From the Vlasov-Poisson system of equations
one may derive, by considering other suitable approximations, the
equations describing the evolution of a self-gravitating fluid. In
this treatment the discrete nature of a particle distribution is
neglected in the dynamical evolution: we will come back on this point
below.

\section{Finite systems}
\label{finite}

We now consider a very simple model to describe the dynamics of an
isolated finite system.  Let us suppose to have an initial  uniform
distribution of $N$ points of identical mass $m$ in a spherical volume
of radius $R_0$ and density
\[
\rho_0 = Nm \frac{3}{4\pi R_0^3} \;.
\]
We suppose that this system is isolated in an infinite space. Let us
consider the limit of a perfect uniform system
\[
\lim_{N \rightarrow \infty, \;\; m \rightarrow 0}  Nm = \mbox{const} \;,
\]
where the fluctuations in the system also go to zero in the same limit.
The mass contained in a sphere of radius $r_0 \le R_0$ is just
\be
M(r_0) = \int_0^{r_0}   4\pi r^2 \rho_0 dr = \frac{4 \pi}{3} r_0^3 \rho_0 \;.
\ee
For reasons of spherical symmetry, 
the force exerted on a particle of unitary mass in the point at distance
$r_0$ from the center of the sphere is due to the matter inside the
shell
\be
F(r_0)= -\frac{G M(r_0)}{r_0^2} \;.
\ee
This force is radial and directed toward the center of the sphere. For
simplicity we suppose that the initial velocities are zero so that the
initial total energy is equal to the initial potential energy and thus
it is negative.  In addition, we make the assumption that different
shells do not overlap during the collapse.  Thus the point initially
at $r_0$ will be attracted by the same mass $M(r_0)$ at all times but
with varying force. The equation of motion can be written as
\be
\label{scm1}
\ddot r(t) =  -\frac{G M(r_0)}{r^2(t)} \;.
\ee
This can be integrated to give
\be
\left(\frac{dr}{dt}\right)^2 = 2GM(r_0) \left(
\frac{1} {r} - \frac{1} {r_0} \right) \;,
\ee
where $G$ is the Newton's constant. 
The solution of this equation can be written in a parametric form
\bea
&& 
r(\xi) = \frac{r_0}{2} (1 + \cos(\xi))
\\ \nonumber 
&&
t(\xi) = \sqrt{\frac{3}{32 \pi G \rho_0}} \left( \xi + \sin(\xi) \right)  \;,
\label{scm2}
\eea
where 
\bea
&&
r(\xi=0)=r_0 \;\;\; \mbox{at} \;\;\; t(\xi=0)=0 
\\ \nonumber 
&& r(\xi=\pi)=0 \;\;\;\mbox{at} 
\; \; t(\xi=\pi)= \sqrt{\frac{3\pi}{32 G \rho_0}}
\equiv \tau_{scm} \;.
\eea
Thus this evolution describes the collapse from the time $t=0$ at
which the system has radius $R_0$ to the time $\tau_{scm}$ at which
the system has zero radius. At longer times the system will re-expand
up to reach the initial configurations, and then will continue these
oscillations periodically. It is interesting to note that in this
model the system never reaches a virialized state (although the total
energy is negative).

\begin{figure}
  \includegraphics*[height=.3\textheight]{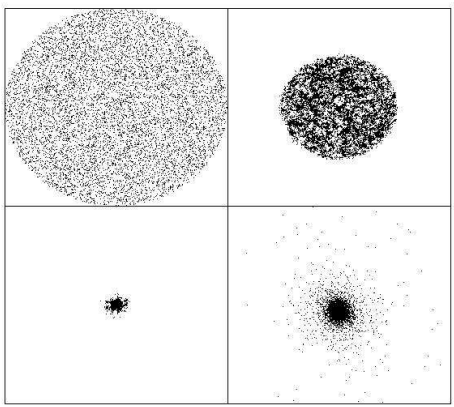} \caption{Different
  phases of the spherical cold collapse (projection of the $x-y$ plane
  of a thin slice around the center of the system, chosen, at
  different times, to have the approximately the same number of points
  inside): top-left initial distribution, top-right at $t=0.8
  \tau_{scm}$, bottom-left at $t=1 \tau_{scm}$ and bottom-right at
  $t=2 \tau_{scm}$. From this latter time on the structure is
  virialized and a forms quasi-stationary state.}
\label{fig1}
\end{figure}

There are some essential aspects of the problem which this 
approach is missing. It is in fact well known through the studies of
computer experiments, that a system of particles is well described by
Eq.\ref{scm2} at times prior to $\tau_{scm}$, while it differs for $t
\rightarrow \tau_{scm}$, when the particle system goes through a phase
of collective relaxation which brings a part of particles to form a
quasi-equilibrium configuration in virial equilibrium. Total collapse
will never occur because small initial inhomogeneities in the system,
always present in any particle distribution, will generate random
motions which will eventually stop the collapse (see Fig.\ref{fig1}).

\begin{figure}
  \includegraphics*[height=.3\textheight]{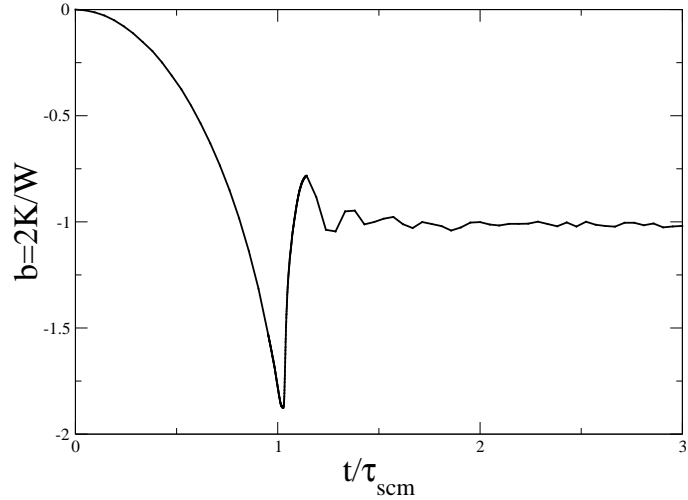} 
\caption{ Virial
  ratio $b=2K/W$, where $K$ is the total kinetic energy of the bound
  particles and $W$ their gravitational potential energy, as a function
  of time: At times $t>\tau_{scm}$ we have that $b\approx -1$ and hence the
  virial theorem is satisfied.  }
\label{fig2}
\end{figure}

In this situation the system forms a state characterized by a dense
central region in virial equilibrium surrounded by a low-density halo
made of particles with positive total energy. The subsequent evolution
is driven by collisions: particles in the high density central region
can undergo to two-body scatterings. The integrated effect of these
collisions is that a part of the particles gain some kinetic energy
while the others end up in a more bounded state.  The long term
consequence of these close encounters is the so-called evaporation,
i.e. that the core becomes more and more bounded and looses little by
little its particles by ejecting them. Thus the asymptotic state
should be made of very few (in principle two) particles orbiting one
around the others, very close so to keep all the potential energy of
the system, and an ideal gas made by the other particles which can
move freely bringing the main part of the kinetic energy of the
system (see Fig.\ref{fig2}).

Lynden-Bell \cite{lyndebell}, who named the collective relaxation
process as ``violent relaxation'', made the first attempt to construct
a theory describing the process of gravitational collapse by
considering a collision-less system.  While the process of violent
relaxation is the main mechanism for relaxation of self-gravitating
systems, in very different situations, it is still not completely
understood in its full details and the theory of Lynden-Bell has shown
various disagreements when compared with the results with N-body
simulations even for the simplest case illustrated above (the same
occurs with other theoretical attempts --- see e.g. \cite{violrelax}).

The violent relaxation process acts on a time scale $\tau_{scm}$ which
is much shorter than, for example, the typical time scale for two-body
collisions $\tau_2$, which sets the time scale for the evaporation of
the core-halo structure mentioned above. Indeed one may show
\cite{saslaw2,binney} that for a 
system of $N$ particles this goes as
\[
\tau_2 \propto \frac{N}{\log(N)} \tau_{scm} \;.
\]
This situation outlines an important difference between systems with
long and short range interactions.  As mentioned, for systems with
short-range forces the relaxation proceeds through
collisions. Self-gravitating systems instead relax through processes
other than the collisional relaxation, such as violent relaxation,
which operate on time scales which are much shorter than the two-body
relaxation time. In fact, the collisional relaxation time for several
astrophysical systems is larger than the age of the universe and for
many years this was considered a paradox (see e.g. \cite{saslaw2}):
how can globular cluster be in virial equilibrium if their relaxation
time (supposed to be given by $\tau_2$) is longer than the age of the
universe ? The answer was simply that the relaxation mechanism was
different from two-body collisions.

Thus the complete theoretical characterization of the simple spherical
collapse of cold particles described above, is an important and still
open problem not only for astrophysics but also for cosmology. Indeed
in both contexts one observes, in numerical simulations, the
formations of core-halo structures (we will come back below on the
cosmological simulations and on the structures formed therein). Thus a
mechanism similar to the violent relaxation of a finite system is
essential also for the virialization of larger and larger clusters of
particles in an infinite system although with some important
differences
\cite{white,morikawa}. As in the finite system, also in the infinite
system the two-body relaxation time scale is much longer than the real
relaxation time of structures. Because the mean field force is due to
fluctuations, the time-scales for the collapse of an over-density of
given size is much longer in the infinite system than in the finite
system case.

The ultimate task of a theory describing the process of violent
relaxation should be the prediction of the quasi-stationary
(virialized) state, which can be generally described by the Vlasov
equation, given the initial conditions in terms of the phase space
density (see e.g. \cite{hansen}). In particular there is a full set of
properties of the virialized state, such as the density profile, the
velocity distribution and profile, and their relations, which seems to
be universal, i.e. appearing from many different initial conditions
(see e.g. \cite{morikawa}) and that are currently unexplained from a
theoretical point of view.

Finally we note that being the two-body relaxation time larger than
$\tau_{scm}$ when the number of particles is large enough, the
collisional effects can usually be neglected for self-gravitating
quasi-equilibrium virial states, which can be treated with an
appropriate non-collisional (i.e. Vlasov) approximation. However in
general the problem consists in the comprehension of the role of
collisions and of the terms related to discreteness (i.e. random
forces, etc.) during the out-of-equilibrium phase, i.e. during
collapse giving rise to a stationary configuration. We will come back
to this point in what follows.

\section{Infinite systems}
\label{infinite}
\begin{figure}
  \includegraphics[height=.3\textheight]{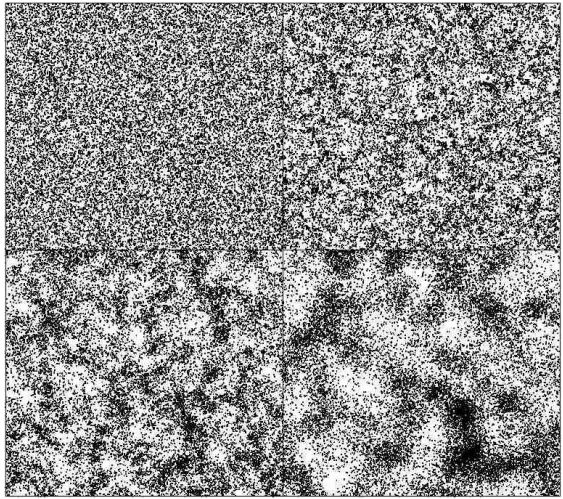} 
\caption{Evolution of the fluctuations and formation of structures
in a simulation (with periodic boundary conditions, representing the
infinite system case) started from cold Poisson initial conditions.
Structures form firstly on small scales (top-left and top-right) and
then propagate to larger and large scales (bottom-left and
bottom-right).}
\label{fig5}
\end{figure}
The problem of the evolution of self-gravitating classical bodies,
initially distributed very uniformly in infinite space, is as old as
Newton. Modern cosmology poses essentially the same problem as the
matter in the universe is now believed to consist predominantly of
almost purely self-gravitating particles --- so called dark matter ---
which is, at early times, indeed very close to uniformly distributed
in the universe, and at densities at which quantum effects are
completely negligible. Despite the age of the problem and the
impressive advances of modern cosmology in recent years, our
understanding of it remains, however, very incomplete.  In its
essentials it is a simple well posed problem of classical statistical
mechanics. 

The standard paradigm for formation of large scale structure in the
universe is based on the growth of small initial density fluctuations
in a homogeneous and isotropic medium. In the currently most popular
cosmological models, a dominant fraction (more than 80 \%) of the
clustering matter in the universe is assumed to be in the form of
microscopic particles which interact essentially only by their
self-gravity.  At the macroscopic scales of interest in cosmology the
evolution of the distribution of this matter is then very well
described by the Vlasov equation coupled with the Poisson equation
\cite{binney,pad_physrep,saslaw1,saslaw2}.

A full solution, either analytical or numerical, of these equations
starting from appropriate initial conditions is not feasible. In
cosmology perturbative approaches to the problem, which treat the very
limited range of low to modest amplitude deviations from uniformity,
have been developed but numerical simulations are essentially the only
instrument beyond this regime and to study the non-linear phase
\footnote{By non-linear objects we mean the structures formed having a
typical density $\rho$ such that the density contrast is $\delta =
(\rho -\rho_0)/\rho_0 \gg 1$, where $\rho_0$ is the average density of
the distribution on  large enough scales. By non-linear clustering we
mean the dynamical evolution leading to the formation of non-linear
objects.}.  N-body simulations solve numerically for the evolution of
a system of $N$ particles interacting purely through gravity, with a
softening at very small scales\footnote{One of the main concerns is
the independency on the softening length of the relevant statistical
quantities}. The number of particles $N$ in the very largest current
simulations is $\sim 10^{10}$ \cite{millenium}, many more than two
decades ago, but still many orders of magnitude fewer than the number
of real dark matter particles ($\sim 10^{80}$ in a comparable volume
for a typical candidate).  While such simulations constitute a very
powerful and essential tool, they lack the valuable guidance which a
fuller analytic understanding of the problem would provide.  The
question inevitably arises of the accuracy with which these
``macro-particles'' trace the desired correlation properties of the
theoretical models. This is the problem of discreteness in
cosmological N-body simulations. It is an issue which is of
considerable importance as cosmology requires ever more precise
predictions for its models for comparison with observations.

\begin{figure}
  \includegraphics[height=.3\textheight]{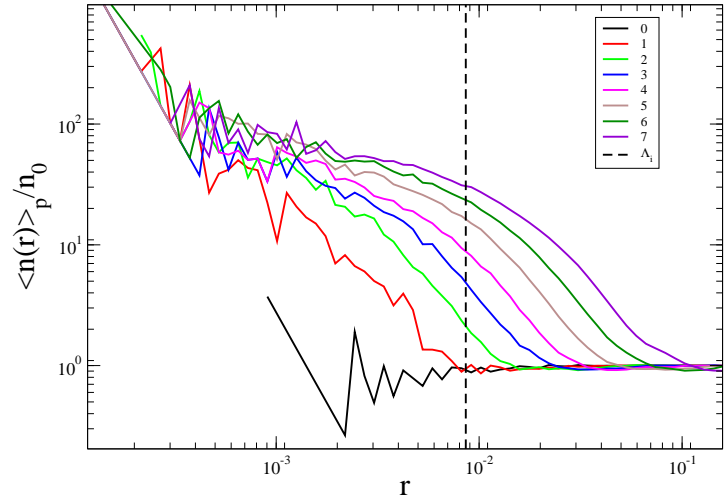} \caption{Evolution
  of the two-point correlation function in a Poisson simulation with periodic
  boundary conditions (representing an infinite system). The temporal
  evolution of this statistical quantity is equivalent to a rescaling
  of the spatial variables. The time indicated in the labels is given
  in units of the dynamical time scale $\tau=\sqrt{4 \pi G
  \rho_0}^{-1}$ of the system. The distance scale $\Lambda_i$ marks 
the initial average distance between nearest particles.}
\label{fig4}
\end{figure}

As already mentioned, the infinite system can never reach a time
independent state (see Fig.\ref{fig5})\footnote{To simulate an
  infinite system one considers a finite volume and infinite replicas
  of it. The force acting on a point is due to all particles inside
  the volume and all replicas, i.e. periodic boundary conditions are
  used.}. However one of the important results from numerical
simulations of such systems in the context of cosmology is that the
system nevertheless reaches a kind of scaling regime, in which the
temporal evolution is equivalent to a rescaling of the spatial
variables \cite{efts88,smith2003}. This spatio-temporal scaling
relation is referred to as ``self-similarity''\footnote{Note that in
  this context the term ``self-similarity'' is used in a completely
  different meaning than usually in statistical physics.  For this
  reason we always use quotes to mark this fact.} and this behavior is
usually observed for the time evolution of the two-point correlation
function or its Fourier transform, the power spectrum (see
Fig.\ref{fig4}). Before describing this evolution in more detail let
us come back to the determination of the gravitational force on an
average particle in a certain distribution and to the approximations
usually employed to study clustering in an infinite system.

\subsection{The force distribution}

Let us consider a uniform particle distribution
\cite{book} in a finite volume $V$.  
The actual value of the gravitational force per unit mass acting on a
fixed ``test'' particle belonging to the system, supposed to be on the
point $\vec{x}$ due to all the other $N$ system particles is given
by
\be  
\label{force1}  
\vec{F}(\vec{x}) = - Gm \sum_{i=1}^N \frac{ \vec{x}  - \vec{x}_i}
{|\vec{x} -\vec{x}_i|^3} 
\ee  
where $m$ is the particle's mass, supposed to be the same for all
particles, and $\vec{x}_i$ the position vector of the $i^{th}$
particle.  The sum in Eq.\ref{force1} is extended to all the particles
other than the test one, which are contained in the system volume $V$
Eventually $V \rightarrow \infty$ and $N \rightarrow \infty$ in such a
way that the average number density $n_0=N/V$ remains constant.  The
actual value of $\vec{F}$ clearly depends on the nature of the local
particle distribution, and hence, it will be in general subjected to
stochastic fluctuations. These fluctuations determine a more or less
broad force probability density function (PDF) $P(\vec{F})$.  Because
of statistical isotropy the direction of $\vec{F}$ is completely
random with equal probability in each direction, i.e. $P(\vec{F})$
depends only on $F=|\vec{F}|$. Thus one may simply consider
$P(\vec{F})d
\vec{F} =4\pi F^2 W(F)dF$, where $W(F)$ is the PDF of the modulus of
the force. 
For reasons of spherical symmetry the average
gravitational force acting on one system's particle and due to the
rest of the system vanishes.  Any force acting on a particle is due to
fluctuations away from exact (i.e. deterministic) spherical symmetry.

Chandrasekhar \cite{chandra_revmodphy} (see also \cite{book,force})
has considered the behavior of the PDF of the Newtonian gravitational
force  arising from a statistically isotropic
(infinite) Poisson distribution of sources.  He showed that applying
the Markov method, it is possible to compute exactly the PDF, known as
the Holtzmark distribution, of the gravitational force acting on a
test particle in the system.

\begin{figure}
  \includegraphics[height=.3\textheight]{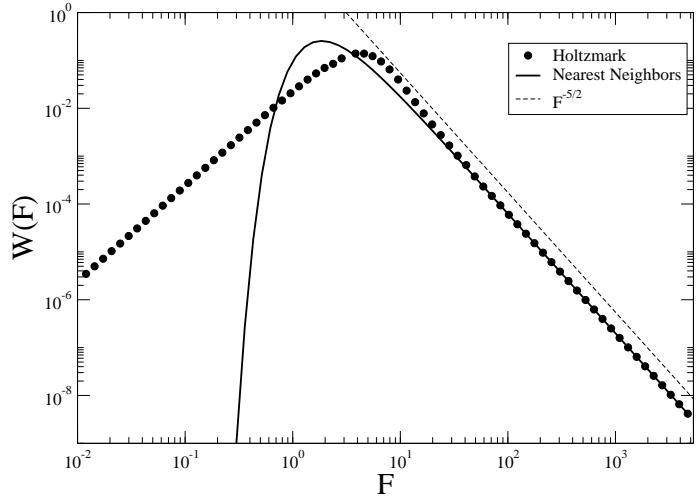} \caption{Holtzmark
  distribution and the PDF inferred if only NNs contribute, i.e.,
  $W_{NN}(F)dF=\omega (r) dr$ being $\omega(r)$ the NN PDF.  The
  agreement is very good in the strong field limit where $W(F) \sim
  F^{-5/2}$. At weak field the PDF due to the NN has a sharp cut-off
  while the full Holtzmark distribution shows a more gentle decrease
  (see discussion in \cite{book,force,earlytimes}).}
\label{fig3}
\end{figure}

A very simple and approximate way to compute this PDF is to consider
only the contribution of the first neighbor: in this way from the
nearest neighbor (NN) probability distribution one may get the PDF of
the force. An interesting point is that the Holtzmark distribution and
the approximated PDF derived by considering the contribution of only
the NN agree very well in the large $F$ region (see Fig.\ref{fig3}). 
 The region where they
differ most is when $F \rightarrow 0$. This is due to the fact that a
weak force can arise only from a more or less symmetric configuration
of particles around the test one in which fluctuations are determined
by many particle effects, and hence the NN approximation
fails. Instead in the strong field limit we may almost neglect the
contribution to the force from far away points, because the main
contribution is due to the limit $r \rightarrow 0$ in the elementary
interaction (i.e. it comes from the NN).  Note that, as in the NN
case, because of the behavior of $W(F)$ for $F\rightarrow \infty$,
$\langle F^2\rangle$ diverges. This is due to the singularity of the
particle-particle gravitational interaction $\sim 1/r^2$ at $r=0$
together with the fact that in the Poisson distribution there is no
explicit positive minimal distance between particles (i.e. no lower
cut-off), as they are permitted to be at an arbitrarily small distance
from one another.

The extension of the Holtzmark distribution to other statistically
homogeneous and isotropic distributions is in general not
straightforward, the main complication being introduced by the presence
of spatial correlations among particles. One may find in
\cite{book,force,masucci,pellegrini} (and references therein) the
derivation of the PDF, under certain assumptions, for particle
distributions with non-trivial correlations.

\subsection{The infinite static case}

In the previous derivation we used an important assumption (see also
\cite{joyce} in this volume), which we now examine in more detail.
The sum defined in Eq.\ref{force1} is not well defined because it is
only conditionally convergent for $V \rightarrow \infty$, i.e.  its
result depends on the order in which the single terms are summed.
Thus the assumption is that the sum has been taken symmetrically with
respect to the point $\vec{x}$. Other prescriptions will in general
lead to different results. To see more clearly this fact  let us
consider the local particle density embedded in a uniform background
of negative mass density so that the microscopic mass density is 
\be
\delta \rho(\vec{x}) = m(n(\vec{x}) -n_0)
\ee
where $n(\vec{x})$ is the local particle density with average density
$n_0$ and the second term represents the uniform background. In this
situation the force on a particle is a well defined stochastic
quantity, i.e.  it does not depend on how the volume $V$ is sent to
infinity \cite{force}. Given that the limit $V \rightarrow \infty$
does not depend on the way the limit is performed, one can choose for
simplicity to take the volume $V$ symmetric with respect to the point
$\vec{x}$ where the force is computed. In such a volume the
contribution to the force from the uniform background vanishes by
symmetry and thus, with this choice of the volume $V$, the force
coincides with the limit of the sum in Eq.\ref{force1}.

While it is possible to show that the presence of an analogous
background with negative mass density comes naturally when the motion
of a particle is described in comoving coordinates in an expanding
universe, in pure Newtonian gravity such a background does not exist
and has to be introduced artificially to regularize the problem (Jean's
swindle). The negative mass density background is equivalent to the 
condition that the force is summed symmetrically (see also 
\cite{force,joyce}).  Note that  
this modification does not necessarily make the gravitational force
well defined in general: whether it is well defined depends on the
nature of the correlations between fluctuations in the density field
on large scales (see discussion in \cite{force}) .

Let us now consider the evolution from an initially cold (i.e. zero
velocity) Poisson distribution. Given that the PDF of the
gravitational force is well approximated by the NN one for large
fields, we can make a simple test: run a gravitational N-body
simulation starting from a cold Poisson distribution (nominally
infinite), with only this component of the force and then compare it
with the situation occurring in the case the full gravity force is
considered \cite{bsl04}. The approximate NN simulation is able to
reproduce the formation of the first structures, made by clusters of
two particles, up to the typical time scale $\tau_{NN}$ for two
particles initially placed at a distance $\ell$, the initial average
distance between NN, to collapse on each other.  This is (again) of the 
order of 
\be
\label{nntime}
\tau_{NN} \sim \sqrt{G m/\ell^3}^{-1} \sim \sqrt{G \rho_0}^{-1} \;,
\ee
where $\rho_0 = m n_0$ and $n_0 \sim \ell^{-3}$ is the number
density. In particular it is very interesting to note that the form of
the two-point correlation function in the non-linear regime, which as
already mentioned will be preserved during time evolution because of
its ``self-similar'' behavior, forms during this NN phase. This is not
all however as the NN truncation will loose an important aspect of the
gravitational evolution, that on large scales, i.e. $r \gg \ell$, 
small-amplitude density fluctuations are growing. This is an effect of
the long-range interactions, of weak amplitude, characterizing the
gravitational dynamics. Let us see briefly how can one model this
phenomenon.

By considering the effect of the uniform negative background, 
the Poisson equation becomes
\be
\nabla \vec{g} = - 4\pi G m (n(\vec{x}) -n_0)
\ee
where $\vec{g}$ is the gravitational field. By considering the
Vlasov-Poisson system of equations, after certain approximations, one
may derive the fluid equations describing the evolution of a
self-gravitating fluid (see e.g. \cite{peebles}). By performing a
perturbation analysis of these equations, for the case of
pressure-less matter around $\rho_0=mn_0$ and $\vec{\upsilon}=\vec{0}$
one finds at first order that the evolution of the density contrast
\[
\delta(\vec{x},t) = \frac{n(\vec{x,t}) -n_0}{n_0}
\]
is described, for the case in which the initial velocity is set equal
to zero, by
\be
\label{linear} 
\delta(\vec{x},t) = \delta(\vec{x},0) \cosh\left( \sqrt{4 \pi G \rho_0} t 
\right) =
\delta(\vec{x},0) \cosh\left(t/\tau \right)
\ee
where $\tau$ is the unique characteristic time scale of the system
which we have already mentioned in Eq.\ref{tau}.  and which is of the
order of the NN collapse time scale given by Eq.\ref{nntime}, i.e. the
fluid dynamics and the NN dynamics are characterized by the same time
scale. This is the only time scale in the systems which will
characterize the evolution of all scales, from $\ell$ to the largest
scales in the system.

Eq.\ref{linear} describes one of the most basic results (see
e.g. \cite{peebles}) about self-gravitating systems, treated in a
fluid limit: that the amplitude of small fluctuations grows
monotonically in time, in a way which is independent of the
scale. This linearized treatment breaks down at any given scale when
the relative fluctuation $\delta$ at the same scale becomes of order
unity, signaling the onset of the ``non-linear'' phase of
gravitational collapse of the mass in regions of the corresponding
size. When the non-linear phase will involve many particles, the
objects formed will have properties similar to the core-halo structure
formed in the collapse of the finite system discussed above. These are
quasi stationary states which naturally emerge from the dynamical
evolution of an infinite self-gravitating system of particles starting
from quasi uniform initial conditions. These are the primary building
blocks in terms of which the non-linear structures observed in
cosmological simulations are described. The physical mechanism
driving the formation of these quasi-equilibrium states is similar to
the violent relaxation process described above, which in this case
occurs in more complicated situations (presence of sub-structures,
etc.) and environments (tidal effects from the boundary conditions,
etc.). Theoretically little is known about the dynamics 
of these objects.

 As discussed in more detail by \cite{joyce} (and see also
 \cite{gravclusl,earlytimes}) the interplay between the small and
 large scales, discrete and fluid dynamics, is an important element
 for the comprehension of the formation of non-linear clustering in an
 infinite system. Large scale fluctuations, i.e. linear theory,
 determine the amplitude of correlations: in this way one expects
 independence on discreteness for the relevant quantities as in the
 fluid limit the discrete scale is sent to zero. On the other hand the
 form of the self-similar two-point correlation function is established in the
 transient phase dominated by two-body interactions, with a slightly
 different time-dependence than fluid linear theory because of the
 effect of discrete fluctuations \cite{plt1,plt2}.  

The considerations of these elements \cite{cg} (see also the
contribution by \cite{joyce} in this volume) lead to the formulation of a
simple model to explain the origin of ``self-similarity'' in the
gravitational clustering of an infinite distribution, where timescales
are dictated by the fluid limit but its non-linear dynamics are
intrinsically discrete. Given that discrete fluctuations characterize
any particle distribution, it is possible to understand why the
two-point correlation function is independent (or very weakly
dependent) on initial conditions.

It has been found \cite{univ} that non-linear clustering which results
in very different simulations is essentially the same, with a
well-defined simple power-law behavior in the two-point
correlations. This in fact a common feature of different sets of
gravitational N-body simulations with different initial particle
configurations, all describing the dynamics of discrete particles with
a small initial velocity dispersion (with or without the space
expansion of cosmological models). This apparently universal behavior
can be understood by the domination of the small scale contribution to
the gravitational force, coming initially from NN particles. In this
perspective the nature of clustering in the non-linear regime has
little to do with the initial fluctuations, or with the space
background being in expansion. Rather it is associated with what is
common to all these simulations: their evolution in the non-linear
regime is dominated by fluctuations on small scales, which are similar
in all cases at the time this clustering develops.

On the other hand is worth noticing that the usual theoretical
modeling (see e.g. \cite{peebles}) considers only a mean-field
approximation (Vlasov limit) and finds non universal behavior of the
two-point correlation function.  In fact, in the cosmological
literature the idea is widely dispersed that the exponents in
non-linear clustering are related to that of the initial correlations
of the small fluctuations in the matter fluid, and even that the
non-linear two-point correlation can be considered an analytic
function of the initial two-point correlations
\cite{pd}. The models used to explain the behavior in the non-linear
regime usually involve both the expansion of the universe, and a
description of the clustering in terms of the evolution of a
continuous fluid. In this perspective the discrete nature of a
particle distribution is neglected. On the other hand discussion of
the small-dynamics at early time presented above already points toward
the fact that a complete characterization of the formation of
non-linear structures it is then required a careful study of the role
of discrete effects in the linear and non-linear regime (see also
\cite{joyce}).

\subsection{The infinite expanding case}

\begin{figure}
  \includegraphics[height=.3\textheight]{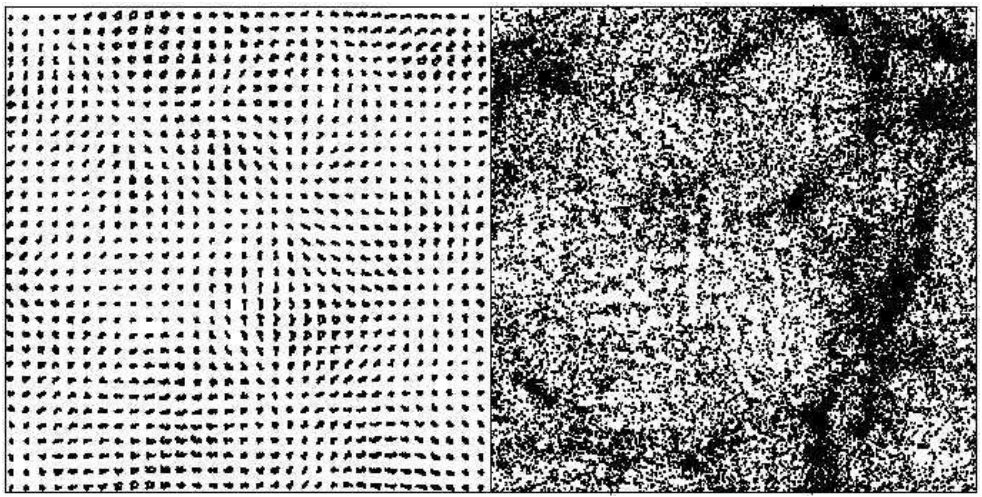} 
\caption{Evolution of the fluctuations and formation of structures
in a simulation (with periodic boundary conditions, representing the
infinite system case) started from a correlated and cold initial
conditions.  Even in this case structures form firstly on small scales
and then propagate to larger and larger scales. The initial conditions
are generated by applying a correlated displacement field to a
pre-initial perfect cubic lattice configuration (see
\cite{bsl02,jbruno}). One may note the appearance of ``filaments'', linking
clusters of particles of different sizes, marking the trace of
long-range correlations in the initial particle distribution. Indeed,
models of primordial density fields in the early universe predict the
presence of long-range weak amplitude correlations on small scales; on
large scales (i.e., $r>100$ Mpc) instead the matter distribution
should have super-homogeneous (or hyper-uniform) features (see
discussion in, e.g.,
\cite{book,glass,slv07}).}
\label{fig6}
\end{figure}

Let us now turn to the more complex case of cosmological structure
formation. In this case one wants to understand gravitational
clustering in an expanding universe. One models this evolution
starting from the Friedmann solution of Einstein's field equations,
where the main assumption is that the matter density field is
considered as a perfectly uniform fluid at all scales
\cite{saslaw2,peebles}. It is then generally assumed that there is a
single length scale characterizing the correlation properties between
density fluctuations, which is of order $100$ Mpc\footnote{We refer 
the interested reader to 
\cite{book,glass,slv07} for a discussion about the correlation
properties of matter density fields in standard cosmological models.},
i.e. much smaller than radius of curvature of the universe (of order
$10^3$ Mpc) and that particles have velocities much smaller than the
velocity of light. In this situation one can treat the problem of
structure formation by assuming the approximation provided with
Newtonian mechanics in an expanding background (see Fig.\ref{fig6}).

The equation of motion of particle in an expanding background is then
\be
\ddot{\vec x}_i + 2 H(t) \vec{x}_i = -\frac{1}{a^3} \sum_{i \ne j} Gm_j 
\frac{\vec{x}_i -\vec{x}_j} {|\vec{x}_i -\vec{x}_j|^3}
\ee
where the dots dots denote derivatives with respect to time,
$\vec{x}_i$ is the comoving position of the $i^{th}$ particle, of mass
$m_i$, related to the physical coordinate by $\vec{r}_i =
a(t)\vec{x}_i$, and where $a(t)$ is the scale factor of the background
cosmology with Hubble constant $H(t) = \dot{a}/a$.  The static case
can be obtained when $a(t)=1$ and thus $H(t) =0$. The expansion rate
$a(t)$ is determined by the type of energy component which dominates
the universe on large scales. For example one may consider a
matter-dominated universe or the case where expansion is dominated by
a cosmological constant (dark energy). In current cosmological models
the $70\%$ of the energy is made of dark energy, $25\%$ of dark matter
and $5 \%$ of ordinary baryonic matter.

In this situation one can develop a perturbation treatment of the
self-gravitating fluid equations in an expanding universe. The results
are similar with the static case discussed above and in a
matter-dominated universe linear theory describes a growing and a
decaying mode, both of them power laws in time rather than exponential
as in Eq.\ref{linear}. Thus in the linear regime at least, it is
possible to map the evolution in a static and in an expanding universe
by taking into account the different density fluctuations growth time
rates.

In the non-linear regime, when cold initial conditions are considered,
one observes the same general characteristics found in the static
background case: a bottom-up aggregation process leading to a
``self-similar'' evolution of the correlation function (in the sense
of Fig.\ref{fig4})\cite{efts88}. The time-rate of growth is power-law
instead of exponential also in the non-linear regime.

The simplest approach in modeling the large scale evolution into
non-linear matter structure is represented by the collapse of a single
over-dense region into a self-gravitating halo via the same spherical
collapse model used to study the collapse of a finite system (see e.g. 
\cite{peacock,coles}). It is interesting to note that the radius of an 
over-dense sphere behaves in the same way as the expansion factor for a
closed universe and therefore one is able to model the growth of a
spherically symmetric density perturbation using the same equations as
classical cosmology, i.e. the Friedmann equations.

The collapse into self-gravitating virialized objects is then a common
feature to finite and infinite systems. In the context of cosmological
N-body simulations the core-halo structures are simply called halos,
by which it is intended the virialized part of these structures. Halos
are ubiquitous in N-body simulations and they show interesting
universal properties (density profiles, velocity distributions),
although different from the those of the finite system described
above. Even in this situation one would like to develop an analytic
treatment to understand the formation of these objects and their
properties. This is still lacking despite the fact that there have
been several attempts for an analytical derivation of the properties
of these halos, and in particular of their density and velocity
profiles (see \cite{morikawa} in this volume).  Standard approaches
are usually based on non-collisional approximations (see
\cite{hansen,white} for a discussion of the subject and for a list of
references).


\section{Conclusions}
\label{conclusions}

The dynamics of infinite self-gravitating systems is a fascinating
theoretical problem of out of equilibrium statistical mechanics,
directly relevant both in the context of cosmology/astrophysics and,
more generally, in the physics of systems with long-range
interactions.  We discussed some of the many problems encountered in
the study of the gravitational clustering in both finite and infinite
systems.  We would like to stress two important open problems.  The
first concerns the extent to which such numerical simulations of a
finite number of particles, reproduce the mean-field/Vlasov limit
which is usually used to describe the evolution from a theoretical
point of view. That is, the theoretical question that arises is about
the validity of this collisionless limit.  Another major question is
that of the understanding of halo structure observed in simulations of
infinite systems: while these show strongly universal characteristics,
their dynamical origin is not yet understood 
from a theoretical point of view.

\begin{theacknowledgments}
  I wish to thank Bruno Marcos, Michael Joyce and Andrea Gabrielli for
  useful comments and fruitful collaborations on the subject. I also
  thank Bill Saslaw for the very many discussions over the years we
  had together on this topic.
\end{theacknowledgments}


\begin{thebibliography}{9}

\bibitem{intro} A. Campa, A. Giansanti, G. Morigi and F. 
Sylos Labini 
 {\it in this volume} 


\bibitem{campa} A. Campa, {\it in this volume}



\bibitem{chavanis} P.H. Chavanis, {\it in this volume} 

\bibitem{lyndebell} D. Lynden-Bell, Mon. Not. R. Astron. Soc.,
{\bf 136}, 101, (1967).  

\bibitem{pad_physrep}
T. Padmanabhan, Physics Reports, {\bf 188}, 285 (1990)


\bibitem{chandra} S. Chandrasekhar, {\it ``Principles of Stellar Dynamics''},
New York: Dover, 1960 

\bibitem{chandra_revmodphy}    S. Chandrasekhar,   
Rev. Mod. Phys.,
{\bf 15}, 1, (1943)  

\bibitem{pad_book} T. Padmanabhan, 
{\it ``Structure formation in the universe''}, 
(Cambridge University Press, Cambridge, 1993)

\bibitem{pad_dtslri}T. Padmanabhan in 
{\it ``Dynamics and Thermodynamics of Systems with Long-Range
  Interaction''},  Dauxois, S. Ruffo, E. Arimondo, M. Wilkens, eds.,  (Springer, Berlin 2002)

\bibitem{saslaw1} W. C. Saslaw, {\it ``Gravitational Physics of Stellar and
Galactic Systems''}, (Cambridge, UK: Cambridge Univ. Press,
1985) 

\bibitem{saslaw2} W. C. Saslaw, 
{\it ``The Distribution of the Galaxies: Gravitational Clustering in
Cosmology''} (Cambridge, UK: Cambridge Univ. Press, 2000).


\bibitem{binney} 
J. Binney and S. Tremaine, {\it ``Galactic Dynamics''} (Princeton
University Press, 1994).


\bibitem{peebles}
P. J. E.,  Peebles,  {\it The Large-Scale Structure of the Universe}
 (Princeton University Press, Princeton New Jersey, 1980)


\bibitem{thierry} T. Baertschiger {\it Ph.D. Thesis} 
(University of Geneve, Geneva, Switzerland, 2004) 


\bibitem{peacock}J.A.  Peacock, 
{\it ``Cosmological physics''} 
(Cambridge University Press, Cambridge, 1999)


\bibitem{morikawa} M. Morikawa, {\it  in this volume}

\bibitem{joyce} M. Joyce,  {\it in this volume}

\bibitem{saslaw} W. C. Saslaw,   {\it in this volume}

\bibitem{binneyknebe}J. Binney and A.Knebe,
 Mon. Not. R. Astron. Soc., {\bf 333}, 378 (2002)


\bibitem{diemond}
J. Diemand, B. Moore, J. Stadel, and S. Kazantzidis,
Mon. Not. R. Astron. Soc.,  {\bf 348}, 977 (2004)

\bibitem{power02} 
C. Power et~al.
Mon. Not. R. Astron. Soc.
{\bf 338}, 14 (2003)


\bibitem{violrelax} 
A. Arad and P.H. Johansson  Mon. Not. R. Astron. Soc. 
{\bf 362},  252 (2005)

\bibitem{white} J.~F. Navarro,
C.~S. Frenk, 
S.~D. White
Astrophys. J., {\bf 490},
493 (1997)


\bibitem{hansen} S. H. Hansen, B. Moore,
New Astron. {\bf 11}  333 (2006)


\bibitem{millenium} 
 V. Springel, et al.,   Nature, {\bf 435},  629 (2005)

\bibitem{efts88}
G. Efstathiou,
C.~S. Frenk,
S.~D.~M. White,
and M. Davis,
Mon. Not. R. Astr. Soc., 
{\bf 235},
715 (1988).

\bibitem{smith2003} 
R.~E. Smith et al., 
Mon. Not. R. Astron. Soc.,
{\bf 341}, 1311 (2003)


\bibitem{book}
 A. Gabrielli, F. Sylos Labini, M. Joyce , L.  Pietronero, {\it
`` Statistical physics for cosmic structures''}, (Springer Verlag,
 Berlin, 2004)



\bibitem{force}  A. Gabrielli, M. Joyce, B. Marcos, F. Sylos Labini,
 T. Baertschiger,  
Phys. Rev. {\bf  E74},  021110  (2006) 

\bibitem{masucci} A. Gabrielli, P. A. Masucci \& F. Sylos Labini,  
Phys. Rev.  {\bf E69},  031110 (2004)



\bibitem{pellegrini}  A. 
Gabrielli, F. Sylos Labini, \& S. Pellegrini, Europhys.Lett., {\bf
46}, 127  (1999)

\bibitem{bsl04} T. Baertschiger \& F. Sylos Labini,  
Phys. Rev.,  {\bf D69}, 123001-1   (2004)

\bibitem{gravclusl} T. Baertschiger, A. Gabrielli, M. Joyce, F. Sylos
Labini, Phys. Rev.{\bf E 75}, 059905 (2007)

\bibitem{earlytimes}T. Baertschiger, M. Joyce, F. Sylos Labini and B. Marcos, 
{\tt arXiv:0711.2219} 

\bibitem{plt1} M. Joyce, B. Marcos, A. Gabrielli, T. Baertschiger,
 F. Sylos Labini, Phys. Rev. Lett., {\bf 95}, 011304 (2005)

\bibitem{plt2} B. Marcos, T. Baertschiger, M. Joyce, A. Gabrielli,
F. Sylos Labini,  Phys. Rev., {\bf D73}, 103507 (2006)


\bibitem{cg}T. Baertschiger, A. Gabrielli, M. Joyce, B. Marcos, 
F. Sylos Labini,  Phys. Rev., {\bf E76}, 011116 (2007)

\bibitem{univ}  F. Sylos Labini, T. Baertschiger \& M. Joyce, 
Europhys.Lett., {\bf 66}, 171 (2004)

\bibitem{glass} A. Gabrielli,  M. Joyce, 
and F. Sylos Labini, Phys. Rev.,  {\bf D65}, 083523 (2002)

\bibitem{slv07} F. Sylos Labini and N.  L. Vasilyev 
Astron. Astrophys., in the press (2007) {\tt arXiv:0710.0224}


\bibitem{bsl02} T. Baertschiger and  F. Sylos Labini,
 Europhys.Lett.,  {\bf 57}, 322 (2002)
  

\bibitem{jbruno}
M. Joyce, D. Levesque, B. Marcos,
Phys.Rev., {\bf D72}, 103509 (2005) 

\bibitem{pd} 
J. Peacock  and S. Dodds, Mon. Not. R. Astron. Soc., {\bf 280}, 
L19 (1996) 

\bibitem{coles} V. Sahni and 
P. Coles,
Physics Reports, {\bf 262},
1 (1995).





\end{thebibliography}
\end{document}